\documentclass[preprint,12pt]{elsarticle}

\usepackage{amssymb}
\usepackage{amsmath}
\usepackage{amssymb}
\usepackage{graphicx}
\usepackage{mathpazo}
\usepackage{wasysym}
\usepackage{epsfig}
\usepackage{amsfonts}
\usepackage{color}
\usepackage{graphics}
\usepackage{epstopdf}
\usepackage{adjustbox}
\usepackage{lipsum}\usepackage{datetime}
\usepackage{color}\usepackage{wasysym}
\usepackage{slashed}
\usepackage{mathrsfs}  
\usepackage{comment}
\usepackage{mathtools}

\journal{Physics of the Dark Universe}

\begin{document}

\begin{frontmatter}



\title{Primordial Gravitational Waves in Generalized Palatini Gravity}


\author[a]{D. Demir}
\address[a]{Faculty of Engineering and Natural Sciences, Sabanc{\i} University, 34956 Tuzla, Istanbul, Türkiye}
\author[b,c]{K. Gabriel}
\address[b]{Department of Physics, Faculty of Science, Cairo University, Giza 12613, Egypt.}
 \address[c]{Center for Fundamental Physics, Zewail City of Science and Technology, 6th of October City, Giza 12578, Egypt.}
\author[c]{A. Kasem}

\author[c]{S. Khalil}


\begin{abstract}
Extended Palatini gravity is the metric-affine gravity theory characterized by zero torsion, nonzero metricity and a quadratic of the antisymmetric Ricci curvature. It reduces dynamically to general relativity plus a geometric Proca field. In this work, we study imprints of the geometric Proca field in the gravitational waves . Our results show that the geometric Proca leaves significant signatures in the gravitational wave signal, and gravitational wave energy density is large enough to be detectable by the next upgrade of the existing GW detectors. Our results, if confirmed observationally, will be an indication that the gravity could well non-Riemannian in nature.
\end{abstract}







\end{frontmatter}


\section{Introduction}
 The detection of gravitational waves (GWs) from binary black hole mergers by LIGO and Virgo collaborations \cite{LIGOScientific:2016aoc} has initiated an era of GW cosmology and opened a new window into the very early stages of the Universe's evolution. General relativity (GR) is currently the most well-established theory of gravity, and it predicts the existence of GWs. However, there are theories beyond the GR that predict GW signals of disparate profiles. Therefore, the detection of GWs can probe alternative theories of gravity and potentially help us understand the nature of gravity on a deeper level. Fortunately, future experimental and observational capabilities will be sensitive enough to probe these theories in depth. As a result, much research attention has been paid to studying GWs phenomena in literature for these scenarios \cite{Barack:2018yly,Heisenberg:2017hwb, Soudi:2018dhv,Garcia-Bellido:2016dkw}.   

 Primordial gravitational waves (PGWs) are relic GWs that were created in the very early universe, shortly after the Big Bang. They predict the observational signatures of tensor perturbation polarization that can be probed in the cosmic microwave background as transverse and traceless parts that carry fundamental degrees of freedom of the GW \cite{BICEP2:2014owc, Keating:2008gp}. PGWs are expected to manifest themselves as a stochastic background signal and thus it is difficult to separate it from various signals of astrophysical origin. However, it is expected that the third-generation GW detectors will be able to discern the signal as they will be sensitive to most compact binary mergers \cite{Sakellariadou:2022tcm}.
 
 PGWs are primary predictions of the vast majority of cosmological models \cite{Christensen:2018iqi, Battye:1996pr, Damour:2004kw}. It is attempted to consider cosmic inflation as one of the common PGW sources, in which they may be generated by quantum fluctuations in the inflaton scalar field. The standard single-field inflationary model predicts a GW energy density of $h^2\Omega_{GW}\sim 10^{-16}$ a scale that is highly unlikely to be probed in the foreseeable future. However, other inflationary models like Starobinsky, chaotic, and hybrid inflation produce much larger PGWs that could be detectable soon enough \cite{Guzzetti:2016mkm}.
 
 GW physics is already fulfilling its promise by providing strong constraints on late-universe which sidelined a large class of scalar-tensor theories. Studying PGWs holds the same promise for theories of the early universe. It is thus reasonable to explore the inflationary mechanisms of modified gravity so that similar constraints can be established. This would help guide the endeavour of searching for alternatives to GW upon the discovery of future signals. The goal of the present work is to probe non-Riemannian geometries in the very early Universe via the observations of the PGWs. More specifically, the goal is to probe the non-metricity vector (gradient of the metric tensor with respect to the affine connection) in Metric-Affine Gravity (MAG) \cite{mag1,Vitagliano2011}. The simplest non-Riemannian extension is the Palatini formulation, which is characterized by a curved metric and Ricci curvature of a general symmetric affine connection (a torsion-free connection independent of the metric and its Levi-Civita connection). This formulation is able to generate the Einstein field equations with no need for extrinsic curvature \cite{york,gh}. With general curvature invariants, it gives GR along with geometrical scalars, vectors and tensors \cite{Demir2012}.
 
 It has been shown that the extension of the Palatini gravity with fundamental scalars like the Higgs field leads to natural inflation \cite{bauer-demir1,bauer-demir2}. Higher-curvature terms were also studied in the Palatini formalism \cite{Vitagliano2011,Demir2020} and their certain effects in astrophysics and cosmology were analysed in \cite{Palatini-f(R)}. One step further from the Palatini formulation is the inclusion of the term quadratic in the anti-symmetric part of the affine curvature. What is important about this inclusion is that it leads dynamically to the GR plus a  purely geometric massive vector theory \cite{Vitagliano2010,Demir2020}. This vector field, a {\it geometric Proca field} as we will call it henceforth, is actually the non-metricity tensor itself \cite{Demir2020,Vitagliano2010}. With a symmetric affine connection (torsion-free), one is left with a special case of non-Riemannian geometries in which non-metricity is the only source of the deviations from the GR. This special case is the Weyl gravity \cite{Weyl0,Weyl1,Weyl2,Vitagliano2013}. The geometric Proca is a direct signature of the Weyl gravity. More specifically, it is significant for metric-incompatible symmetric connections (torsion-free).  It is not introduced put by hand nor originating from gauge theories. In fact, it is a geometrical massive vector field that characterizes the Weyl nature of the geometry \cite{Weyl1,Weyl2}. It has been studied as vector dark matter in \cite{Demir2020}. Its couplings to fermions (quarks and leptons) were explored in \cite{dp-yeni} in regard to the black hole horizon in the presence of the Proca field (see also \cite{uzbek}). It is important to keep in mind that geometric tensor fields come from not only the non-metricity tensor (metric-incompatible symmetric connection) but also the torsion tensor (metric-compatible metric antisymmetric connection) \cite{Kibble:1961ba, Hehl:1976kj, Kasem:2020ddi, Kasem:2020wsp,Iosifidis:2019dua}. The model we are discussing is therefore a special case in that only metric-incompatible connections are considered. 

In analyzing the PGWs from the GR plus geometric Proca setup, we describe the degrees of freedom of the connection (with respect to the Levi-Civita connection) using a set of independent parameters similar to independent components of a massive vector field governed by the Proca equation \cite{Demir2020,dp-yeni}. Our main result is that this geometric Proca field leaves detectable signatures on PGWs, and they form thus a new source of GWs coming from not the matter sector but the geometrical sector. The Proca field is a perturbation on the FRW background generated by the cosmological constant. Our reveals show the dependence of the power spectrum on the initial configuration of the geometric Proca field initial conditions and show that the sensitivity of the GW energy density to the Proca mass is pronounced at large mass values. Our analysis also shows that the GW energy density can be of order $10^{-9}$ -- a value that could be reached by future experiments like the advanced Virgo and the Einstein Telescope.

The paper is structured as follows.  In Sec. \ref{model}, we discuss generalized Palatini gravity and show its reduction to the GR plus the aforementioned geometric Proca field (the canonical massive vector, $Y_\mu$).  In Sec. \ref{PGW}, we analyze tensor perturbations and derive the power spectrum and GW energy density $\Omega_{GWs}$. In Sec. \ref{numerical}, we perform a detailed numerical analysis and put constraints on the field mass, highlight orders of magnitude of the expected GW intensity, and discuss their detection prospects in the near future experiments. Finally, we conclude the work in Sec. \ref{conclusion}. 

\section{\label{model}
Generalized Palatini Gravity}
 
The standard Einstein-Hilbert (EH) theory of GR is based on Riemannian geometry, in which the linear connection is symmetric and  compatible with the metric (satisfying the condition $\nabla_\alpha g_{\mu\nu}=0$). That connection, the Levi-Civita connection, is given by
 \begin{equation}
 	{}^g\Gamma^\lambda_{\mu\nu} = \frac{1}{2} g^{\lambda \rho} \left( \partial_\mu g_{\nu\rho} + \partial_{\nu} g_{\rho\mu}-\partial_\rho g_{\mu\nu}\right).
\label{levi-civta}
\end{equation}
The EH action $\int d^4 x (\frac{1}{2\kappa} R(g)+ {\mathscr{L}}_{m})\sqrt{-g} $ leads to the Einstein field equations 
\begin{eqnarray}
    G_{\mu\nu}(\Gamma)&=&8 \pi G_N T_{\mu\nu}
\label{HE FE}
\end{eqnarray}
upon variation with respect to the metric $g_{\mu\nu}$. Here, $G_{\mu\nu}(\Gamma)\equiv R_{\mu\nu}-\frac{1}{2}Rg_{\mu\nu}$ is the Einstein tensor, and $T_{\mu\nu}\equiv\frac{-2}{\sqrt{-g}}\frac{\partial(\sqrt{-g}\mathscr{L}_{m})}{\partial g^{\mu\nu}}$ is the energy-momentum tensor of matter.  In Palatini formalism, metric and connection are treated as independent variables, and variation of EH action is carried out with respect to both the 10-component metric and the $4^3$-components of the affine connection. Upon variation, $4^2(4-1)/2$ of which vanish rendering the connection symmetric and the remaining 40 components reproduce the compatibility condition (\ref{levi-civta}) as a consequence of the equations of motion (dynamics) rather than a priori assumption \cite{Misner:1973prb}.

Relaxing the assumption of a symmetric connection leads to a more general manifold (Einstein-Cartan) where the antisymmetric part of the connection is encoded in the torsion tensor \cite{Kibble:1961ba, Hehl:1976kj, Kasem:2020ddi, Kasem:2020wsp} $S^\lambda_{\mu \nu}=\Gamma^\lambda_{\mu \nu}-\Gamma^\lambda_{\nu \mu}$. On the other hand, relaxing the nonmetricity condition allows for a more general framework for modifying gravity and has led to the development of a number of interesting models where the metric (unlike GR) is not the only geometro-dynamical degree of freedom. Indeed, as we will show, by introducing a non-metricity tensor we can add new degrees of freedom to the theory, which can couple to both the metric and the affine connection. This results in a class of models known as \textit{generalized Palatini gravity}, which has been the subject of extensive research in recent years. These models have been found to have significant cosmological and astrophysical implications.

Additional degrees of freedom are encoded in the non-metricity tensor $Q_\lambda^{\mu \nu}=\nabla_\lambda g^{\mu\nu}$. Gravitational theories on such manifolds are classified as metric-affine-gravity theories (MAG) \cite{mag1,Iosifidis:2019dua}. MAG is based on a metric tensor $g_{\mu\nu}$ and a symmetric affine connection $\Gamma^\lambda_{\mu\nu}=\Gamma^\lambda_{\nu\mu}$ (having nothing to do with the metrical Levi-Civita connection in GR). In general, such connections differ from the Levi-Civita connection only by the contributions of the non-metricity tensor. In explicit terms \cite{dp-yeni, Vitagliano2010}
\begin{align}\Gamma^\lambda_{\mu\nu} - {}^g\Gamma^\lambda_{\mu\nu} =  \frac{1}{2} g^{\lambda \rho} ( Q_{\mu \nu \rho } + Q_{\nu \mu \rho } - Q_{\rho \mu \nu} )
\label{fark-connection}
\end{align}
where $Q_{\lambda \mu \nu} = - {}^\Gamma \nabla_{\lambda} g_{\mu \nu}$ being the non-metricity tensor. The Levi-Civita connection sets the covariant derivative $\nabla_\mu$ such that $\nabla_\alpha g_{\mu\nu} = 0$ and sets also the metrical Ricci curvature ${R}_{\mu\nu}\left({}^g\Gamma\right)$, which contracts to give the metrical scalar curvature  $R(g) \equiv g^{\mu\nu}{R}_{\mu\nu}\left({}^g\Gamma\right)$.  The affine connection $\Gamma^\lambda_{\mu\nu}$, on the other hand, is independent of the metrical connection ${}^g\Gamma^\lambda_{\mu\nu}$, defines the covariant derivative  ${}^\Gamma \nabla_\mu$, and sets the affine Riemann curvature
\begin{eqnarray}
\label{affine-Riemann}
{\mathbb{R}}^\mu_{\alpha\nu\beta}\left(\Gamma\right) = \partial_\nu \Gamma^\mu_{\beta\alpha} - \partial_\beta \Gamma^\mu_{\nu\alpha} + \Gamma^\mu_{\nu\lambda} \Gamma^\lambda_{\beta\alpha} -\Gamma^\mu_{\beta\lambda} \Gamma^\lambda_{\nu\alpha}
\end{eqnarray}
with ${\mathbb{R}}^\mu_{\alpha\nu\beta}\left(\Gamma\right)=-{\mathbb{R}}^\mu_{\alpha\beta\nu}\left(\Gamma\right)$. Its contractions lead to two distinct Ricci curvatures: the usual Ricci curvature  ${\mathbb{R}}_{\mu\nu}\left(\Gamma\right) = {\mathbb{R}}^\lambda_{\mu\lambda\nu}\left(\Gamma\right)$ and an anti-symmetric Ricci curvature ${\overline{\mathbb{R}}}_{\mu\nu}\left(\Gamma\right) = {\mathbb{R}}^\lambda_{\lambda\mu\nu}\left(\Gamma\right)$. The latter is actually the antisymmetric part of the former 
\begin{eqnarray}
{\overline{\mathbb{R}}}_{\mu\nu}\left(\Gamma\right) = {\mathbb{R}}_{[\mu\nu]} \left(\Gamma\right)  = \partial_\mu \Gamma^\rho_{\rho \nu} - \partial_\nu \Gamma^\rho_{\rho \mu}
\label{asym-curvature}
\end{eqnarray}
which identically vanishes when the affine connection $\Gamma^\lambda_{\mu\nu}$ is replaced with the metrical one ${}^g\Gamma^\lambda_{\mu\nu}$. The total affine Ricci curvature gives the affine scalar curvature $R(g,\Gamma)\equiv g^{\mu\nu}{\mathbb{R}}_{\mu\nu}\left(\Gamma\right)$ \cite{bauer-demir1, Demir2012}. Palatini theories involving the quadratic ${\overline{\mathbb{R}}}_{\mu\nu}{\overline{\mathbb{R}}}^{\mu\nu}$ of the antisymmetric curvature ${\overline{\mathbb{R}}}_{\mu\nu}\left(\Gamma\right)$ in (\ref{asym-curvature}) has been of physical interest in literature. Recent analyses suggested a model in this framework that is ghost-free and free of classical and quantum instabilities, the action is defined as \cite{Demir2020,dp-yeni,Vitagliano2010}
\begin{eqnarray}
S[g,\Gamma]=\int d^4x \sqrt{-g}\left\{
\frac{M^2}{2} {R}\left(g\right) + \frac{{\overline{M}}^2}{2} {\mathbb{R}}\left(g,\Gamma\right)  + \zeta {\overline{\mathbb{R}}}_{\mu\nu}\left(\Gamma\right) {\overline{\mathbb{R}}}^{\mu\nu}\left(\Gamma\right) - V_0
+{\mathscr{L}}_{m}({}^g\Gamma,\psi)\right\}
\label{mag-action}
\end{eqnarray}
the action is composed of physically distinct parts. The first part proportional to $M^2$ would be the usual Einstein-Hilbert action if $M$ were equal to the Planck mass $M_{Pl}$. The second term proportional to ${\overline{M}}^2$ is the standard Palatini action \cite{Demir2012,bauer-demir1}, which leads to the  Einstein field equations with no need to extrinsic curvature \cite{york,gh}. The third term proportional to $\zeta$ was considered in both \cite{Vitagliano2010} and \cite{Demir2020}. The fourth term $V_0$ is the vacuum energy density. The matter Lagrangian ${\mathscr{L}}_{m}({}^g\Gamma,\psi)$ governs dynamics of the matter fields $\psi$ (matter sectors involving  $\Gamma^\lambda_{\mu\nu}$ (not  ${}^g\Gamma^\lambda_{\mu\nu}$) have been analyzed in \cite{Demir2020}). Our setup differs from the so-called metric-Palatini setup  \cite{harko2012,Capozziello2013a,Capozziello2015} by the third term proportional to $\zeta$ (and dropping of the higher powers of ${R}_{\mu\nu}\left({}^g\Gamma\right)$ and ${\mathbb{R}}_{\mu\nu}\left(\Gamma\right)$ to evade gravitational ghosts).

In analyzing the action (\ref{mag-action}), it proves useful to define the non-metricity vector $Q_\mu = Q _{\mu \nu}^{\nu}/4$ as it turns out that upon variation all 40 components can be written in terms of this quantity. In fact, the generalized Palatini action (\ref{mag-action}) written in terms of the independent connection reduces to \cite{Vitagliano2010,Demir2020}
\begin{align}
S[g,Y,\psi] &= \int d^4 x \sqrt{-g} \Bigg \{\frac{1}{16 \pi G_N} R(g) -\frac{\Lambda}{8\pi G_N}
- \frac{1}{4} Y_{\mu \nu} Y^{\mu \nu} - \frac{1}{2} M_Y^2 Y_{\mu} Y^{\mu} + {\mathscr{L}}_{m} (g,{}^g \Gamma,\psi) \Bigg \}
\label{action-reduced}
\end{align}
in the notation of \cite{Demir2020}. In this action, $\Lambda = 8 \pi G_N V_0$ is the cosmological constant, $Y_\mu=2 \sqrt{\zeta} Q_\mu$ is the canonical geometric vector field, $G_N=8\pi/(M^2 + \overline{M}^2)$ is Newton's gravitational constant, and
\begin{eqnarray}
M_Y^2 = \frac{3 \overline{M}^2 }{2 \zeta}\label{Ymass}
\end{eqnarray}
is the squared mass of the $Y_\mu$ (a geometric Proca field). As revealed by (\ref{action-reduced}), the Einstein-Geometric Proca system in (\ref{action-reduced}) involves two degrees of freedom: the metric tensor $g_{\mu\nu}$ and the Proca field $Y_\mu$ -- a purely geometrical field set by the non-metricity vector. 

The action (\ref{action-reduced}) gets extremized against variations in metric provided that the Einstein field equations hold
\begin{align}
G_{\mu \nu} + \Lambda g_{\mu \nu} = 8 \pi G_N \left(T^{(Y)}_{\mu\nu} + T^{(matter)}_{\mu\nu}\right) 
\label{einstein-eq}
\end{align}
where 
\begin{eqnarray}
T^{(Y)}_{\mu\nu}&=&{Y}_{\mu \alpha} {Y}_{\nu}^{\alpha}   - \frac{1}{4} {Y}_{\alpha \beta} {Y}^{\alpha \beta} g_{\mu \nu} + M_Y^2 {Y}_{\mu} {Y}_{\nu}\label{Tmunu}\\
Y_{\alpha\beta}&=&\partial_\alpha A_\beta-\partial_\beta A_\alpha
\end{eqnarray}
is the energy-momentum tensors the geometric Proca $Y_{\mu}$. By a similar analysis, the action (\ref{action-reduced}) remains stationary against variations in ${Y}_\mu$ provided that its Proca equation holds 
\begin{align}
\nabla_\mu {Y}^{\mu \nu} - M^2_Y {Y}^\nu = 0 
\label{eom-Y}
\end{align}
as a free massive vector field. 

In what follows, we shall treat the geometric Proca ${Y}_\mu$ as a test particle (perturbation) on the cosmological background set by the cosmological constant $\Lambda$. In fact, the metric
\begin{eqnarray}
ds^2 = -dt^2 + a^2(t)\delta_{i j} dx^i dx^j
\label{metric-00}
\end{eqnarray}
in the cosmological time $t$. The zeroth order of Einstein field equations (\ref{einstein-eq}) of a de Sitter universe ($G_{\mu\nu}^{(0)}+\Lambda g_{\mu\nu}^{(0)}=0$) yields the standard Friedmann equations which can be readily solved for the scale factor
\begin{eqnarray}
a(t) = a(t_0) e^{\sqrt{\frac{\Lambda}{3}}t} 
\label{scale-factor}
\end{eqnarray}
revealing the exponential growth of the overall scale of the universe with cosmological time. For the Einstein-geometric Proca system in (\ref{action-reduced}), the solution of the scale factor in (\ref{scale-factor}) sets the background geometry, in which $Y_{\mu}(t)$ evolves as a first-order perturbation. In this case, the massive vector field influences only the higher orders of perturbation theory \cite{Fernandes:2022con}.

\section{Cosmic Perturbations Induced Primordial Gravitational Waves}\label{PGW}

We consider the conformal FRW metric perturbed by small perturbation. In the longitudinal gauge, up to the second order, this usual setup can be written as \cite{Mukhanov:2005sc, Acquaviva:2002ud, Baumann:2009ds}
\begin{eqnarray}
g_{00}&=&-a^2(\eta)(1+2\psi^{(1)}+\psi^{(2)})\\
g_{0i}&=&0\\
g_{ij}&=&a^2(\eta)\Big[ (1-2\phi^{(1)}-\phi^{(2)})\delta_{ij}+\frac{1}{2}\big(\partial_i\chi_j^{(2)}+\partial_j\chi_i^{(2)}+h_{ij}^{(2)}\big)\Big].
\end{eqnarray}

The background solution (\ref{scale-factor}) in conformal time takes the form $a(\eta) =-\frac{1}{H\eta}$. Having set the background geometry, we start exploring Einstein field equations in the first order. In this regard, we take the geometric Proca field as a first-order effect so that its gravitational backreaction is second order (the Proca energy-momentum tensor in (\ref{Tmunu}) is second order namely $T_{\mu\nu}^{(1)}=0$) so that first-order Einstein field equations take the form $G_{\mu\nu}^{(1)}=0$.
Transverse components of Einstein equations lead to $\psi^{(1)}=\phi^{(1)}$, the $(0-0)$-component and the $(0-i)$-components imply that perturbations also vanish at the first order: $\psi^{(1)}=\phi^{(1)}=0$. Moreover, the equation of motion of the second-order scalar perturbations $\psi^{(2)}, \phi^{(2)}$ are non-dynamical as well. Hence, in the uniform limit of the geometric Proca field the other modes drop out and we end up with $h_+$ and $h_\times$ only, where one note that in more general Proca field theories additional polarization scalar modes are generated \cite{DeFelice:2016uil}. This is to be contrasted with the treatment of Proca field as a background field where its effects  include also other modes \cite{Adshead:2017hnc, Armendariz-Picon:2004say, Maleknejad:2011jw, DeFelice:2016uil}.

Before we turn to examine the tensor perturbations, we construct an explicit solution of the Proca equation (\ref{eom-Y}) as a first-order perturbation:
\begin{eqnarray}
Y_0^{(1)}(\eta)&=&0\\
Y_i^{(1)}(\eta)&=& C_{i+} \eta^{\frac{1}{2}+\sqrt{1+\frac{m^2}{H^2}}}+C_{i-} \eta^{\frac{1}{2}-\sqrt{1+\frac{m^2}{H^2}}},
\label{Qsol}
\end{eqnarray}
in which $C_{i+}$ and $C_{i-}$ are six constants of integration. The equations governing evolution of tensor perturbations are obtained by acting with the projection operator $\Pi_{ij}^{lm}$ onto transverse modes \cite{Guzzetti:2016mkm, Baumann:2007zm}. This projection eliminates the scalar and vector parts of the second-order Einstein field equations as
\begin{eqnarray}
\Pi_{ij}^{lm} G_{lm}^{(2)} = 8 \pi G_N \Pi_{ij}^{lm}T_{lm}^{(2)}
\label{GW}
\end{eqnarray}
and leads then to the following wave equation 
\begin{eqnarray}
h^{\prime\prime}{}_{ij}+2H h^{\prime}{}_{ij}-\nabla^2 h_{ij}&=&-4\Pi{_{ij}{}^{lm}} \mathcal{S}_{lm}\label{eomft}\\
\mathcal{S}_{lm}(\eta) &\equiv& 8 \pi G_N\left(m^2Y_lY_m-\frac{1}{a^2} Y^{\prime}{}_l Y^{\prime}{}_m\right)\nonumber \\
&-&8 \pi G_N\Big(m^2Y_kY_k-\frac{1}{a^2} Y^{\prime}{}_k Y^{\prime}{}_k\Big)\delta_{lm}
\end{eqnarray}
sourced by the geometric Proca field $Y_\mu$. Expanding $h_{ij}$ into standard polarization tensors $\left(e_{ij}^{(+)}, e_{ij}^{(\times)} \right)$, and Fourier-transforming from ($\eta,{\textbf{x}}$) to ($\eta,{\textbf{k}}$) space \cite{Racco:2018iso}, one gets
\begin{eqnarray}
h^{\prime\prime}{}^{(\lambda)}(\textbf{k},\eta)+2H h^{\prime}{}^{(\lambda)}(\textbf{k},\eta)+k^2 h^{(\lambda)}(\textbf{k},\eta)=\mathcal{S}^{(\lambda)}(\textbf{k},\eta),
\label{GWploar}
\end{eqnarray}
in which $\mathcal{S}^{(\lambda)}(\textbf{k},\eta)\equiv -4e^{(\lambda)}{}^{lm}(\textbf{k})S_{lm}(\eta)\delta^3(\textbf{k})$ and $\lambda$ stands for the polarization modes $\lambda= +$ and $\lambda=\times$. We now introduce a Green function
\begin{eqnarray}
G(\textbf{k},\eta;\tilde{\eta})&=&\frac{1}{\sqrt{H^2-k^2}}\sinh(\sqrt{H^2-k^2}(\eta-\tilde{\eta}))
\label{Green}
\end{eqnarray}
to  solve the equation (\ref{GWploar}) as 
\begin{eqnarray}
h(\textbf{k},\eta)&=&\frac{1}{a(\eta)}\int^{\eta}_{\eta_a} d\tilde{\eta} \ G(\textbf{k},\eta;\tilde{\eta})\ a(\tilde{\eta})\ \mathcal{S}(\textbf{k},\tilde{\eta})\,.
\end{eqnarray}
As usual, the power spectrum of the GWs is defined via the two-point function  
\begin{eqnarray}
\langle h(\textbf{k}_1,\eta)h(\textbf{k}_2,\eta)\rangle &=&\frac{2\pi^2}{k^3}\delta(\textbf{k}_1-\textbf{k}_2)\mathcal{P}_h(k,\eta)
\end{eqnarray}
in which the power density takes the form 
\begin{eqnarray}
\mathcal{P}_h(k,\eta)&=&-16 H^2 \eta^2\int_{\eta_a}^\eta \int_{\eta_a}^\eta \frac{1}{\sqrt{H^2-k_1{^2}}}\sinh(\sqrt{H^2-k_1{^2}}(\eta-\tilde{\eta_1}))\frac{1}{H \tilde{\eta}_1}  \Big[m^2 Y_i(\tilde{\eta}_1)Y_j(\tilde{\eta}_1)\nonumber\\
&-&\frac{1}{H \tilde{\eta}_1}   Y^{\prime}_i(\tilde{\eta}_1)Y^{\prime}_j(\tilde{\eta}_1)\Big]+\frac{1}{\sqrt{H^2-k_2{^2}}}\sinh(\sqrt{H^2-k_2{^2}}(\eta-\tilde{\eta_2})) \frac{1}{H \tilde{\eta}_1}  \Big[m^2Y_l(\tilde{\eta}_2)Y_m(\tilde{\eta}_2)\nonumber\\
&-&\frac{1}{H \tilde{\eta}_2}   Y^{\prime}_l(\tilde{\eta}_2)Y^{\prime}_m(\tilde{\eta}_2)\Big]e^{ij}e^{lm} d\tilde{\eta_1} d\tilde{\eta_2}\label{PS}    
\end{eqnarray}
after a direct calculation of the two-point function
\begin{small}
\begin{eqnarray}
\langle h(\textbf{k}_1,\eta)h(\textbf{k}_2,\eta)\rangle &=&
\frac{1}{a^2(\eta)}\int_{\eta_a}^{\eta}\int_{\eta_a}^{\eta} d\tilde{\eta}_1 d\tilde{\eta}_2  a(\tilde{\eta}_1) a(\tilde{\eta}_2) G(\bold{k},\eta;\tilde{\eta}_1)G(\bold{k},\eta;\tilde{\eta}_2)\nonumber \\
&&\left<\mathcal{S}(\bold{k}_1,\tilde{\eta}_1)\mathcal{S}(\bold{k}_2,\tilde{\eta}_2) \right>
\end{eqnarray}
\end{small}
for the geometric Proca perturbation under concern. Then, the energy density per logarithmic interval in $k$  is defined in terms of the tensor power spectrum as \cite{Racco:2018iso,Maggiore:1999vm, Kohri:2018awv}
 \begin{equation}
     \rho_{GW}(k,\eta)\equiv \frac{1}{64 \pi G_N}\left(\frac{k}{a}\right)^2 \overline{\mathcal{P}_h(k,\eta)},
 \end{equation}
in which the overline stands for time-average over the oscillations. As a result, the egergy density of GWs, denoted as $\Omega_{GW}$, takes the form  
\begin{eqnarray}
\Omega_{GW}(k,\eta)\equiv \frac{\rho_{GW}(k,\eta)}{\rho_{cr}(\eta)}=\frac{1}{24}\left(\frac{k}{H}\right)^2 \overline{\mathcal{P}_h(k,\eta)}\,.
\label{Omega}
\end{eqnarray}
In the next section, we shall perform detailed numerical analyses of $\mathcal{P}_h(k,\eta)$ and $\Omega_{GW}(k,\eta)$ as functions of the Proca sector parameters. 

\section{Numerical Analysis \label{numerical}}

In this section, we perform a numerical analysis of the tensor power spectrum generated by the geometric Proca vector field $Y_i$, and the associated energy density of GWs. As previously stated, the scalar and vector power spectra vanish identically in this class of models because the perturbations of the scalar and vector components of the metric do not evolve dynamically and vanish quickly. It is worth emphasizing that the scalar perturbations lead to variations in the strength of the gravitational field, which cause matter to clump together in some regions and become less dense in others, as observed in the CMB through small temperature fluctuations. It thus is clear that scalar perturbations are a necessary component in the evolution of the Universe, and we assume that they are derived from variations in the matter density (like a perfect fluid or scalar field). Vector perturbations, on the other hand, refer to perturbations in the velocity of matter in the Universe. They inherently imply a preferred direction for the collective motion of the matter. There is no evidence for such a preferred direction in the CMB so it is  fortunate that these perturbations decay quickly.

The power spectrum integral in (\ref{PS}) is too complicated to perform analytically. So, henceforth, we will continue with numerical computations for specific values of the model parameters (Proca sector). As we expect $R^2$ effects to be pronounced at high energies close to the Planck scale, we choose a high-scale inflation $H^\star=10^{14}\ {\rm GeV}$ compatible with the observational upper bound \cite{BICEP2:2015xme, CMB-S4:2016ple}. All model parameters are assumed at this scale unless otherwise stated.

\begin{figure}[h!]
\begin{center}    
\includegraphics[scale=0.6]{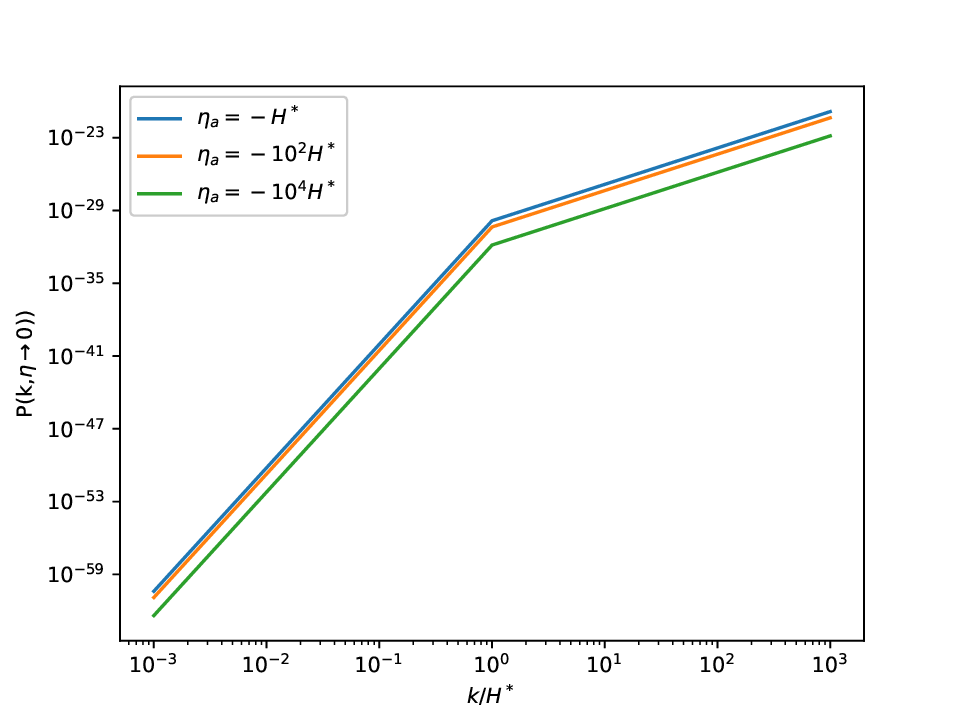}
\caption{The tensor power spectrum $\mathcal{P}_h(k,\eta)$ as a function of $k/H^\star$ near the end of inflation epoch  ($\eta\rightarrow 0$) for different initial values  of $\eta_a$ ($\eta=\eta_a$). In general, smaller the $|\eta_a|$ slightly larger the power spectrum.}
\label{kmodes1}
\end{center}
\end{figure}
Depicted in FIG. \ref{kmodes1} is the variation of the tensor power spectrum $\mathcal{P}_h(k,\eta)$ as a function of the ratio $k/H^\star$ for $M_Y=H^\star$ and the initial $\eta$ values $\eta_a=- H^\star, -10^2 H^\star$ and $-10^4 H^\star$. As this figure suggests, subhorizon modes $k>H^\star$ dominate the power spectrum. This boundary is marked by a change of the slope on the logarithmic scale at $k=H^\star$. This behaviour is suggested by the propagator (\ref{Green}) which changes from decaying to sinusoidal. Physically, this implies that subhorizon modes never exit the horizon during inflation (which enhances their contribution). This result suggests that stochastic gravitational waves background (SGWB) signal could be detected by ground-based interferometers as these devices have an enhanced sensitivity to higher frequencies \cite{Caprini_2018}. It is worth noting that  
the power spectrum decreases at larger $|\eta_a|$ (corresponding to a longer inflationary period). The reason for this decrease is the distribution of the energy of fluctuations over a longer and longer time period. As discussed below, unlike the power spectrum, energy density $\Omega_{GW}(k,\eta)$ grows with large $|\eta_a|$.

\begin{figure}[h!]
\begin{center} 
\includegraphics[scale=0.6]{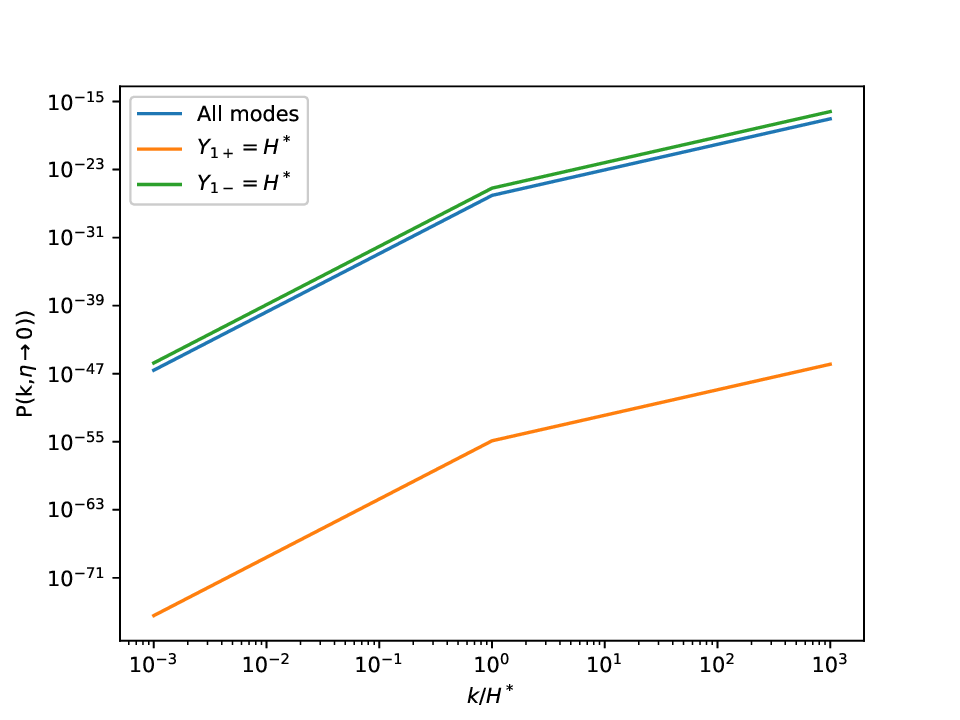}
\caption{The tensor power spectrum $\mathcal{P}_h(k,\eta)$ as function of $k/H^\star$ near the end of inflation epoch  ($\eta\rightarrow 0$) for different initial field values  of the Proca field $Y_i^{(1)}(\eta_a)$. The dependence is seen to be quite strong since $Y_{1+}$ curve lies below $Y_{1-}$ curve by some 24 orders of magnitude.}
\label{kmodes2}
\end{center}
\end{figure}

\begin{figure}[h!]
\begin{center} 
\includegraphics[scale=0.6]{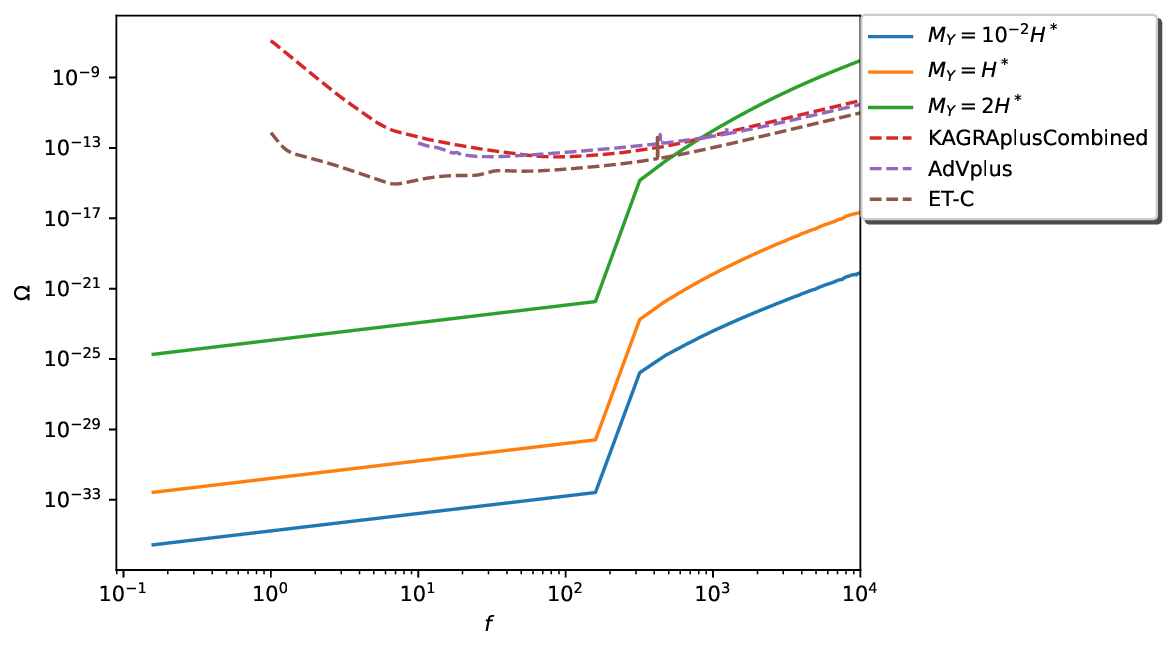}
\caption{The GW energy density $\Omega_{GW}(k,\eta)$ defined in (\ref{Omega}) as a function of the GW frequency $f$ and the geometric Proca mass $M_Y$. Superimposed on the plots are the sensitivity curves for the KAGRA upgrade \cite{Kagra37}, Advanced Virgo \cite{LIGODOC87}, and the Einstein Telescope \cite{ET Sensitivity}. Future GW detectors can detect GW from geometric Proca of mass $M_Y \gtrsim 2 H^\star$.}
\label{OmegaPlot}
\end{center}
\end{figure}

Shown in FIG. \ref{kmodes2} is the dependence of the power spectrum on the initial field values defined in (\ref{Qsol}). It is evident that $Y_{i-}$ modes dominate the power spectrum because these modes grow rapidly towards the end of inflation due to their negative-power-law dependencies on $H$. To put this into perspective, we can compare it to the power spectrum from a pure de Sitter universe. For $k\ll \frac{1}{\mid\eta_i\mid}$ \cite{Mukhanov:2005sc}

\begin{eqnarray}
    P(k,\eta)=\frac{8H^2}{\pi}(1+(k\eta)^2)
\end{eqnarray}
so $P(k,\eta\rightarrow 0)\approx 10^{-9}$ and the Proca field apparently suppresses GW production. However we are yet to investigate the exponent of (\ref{Qsol}) by varying the Proca mass which is a parameter of physical importance being related to fundamental coupling constant (\ref{Ymass}). To reveal its role on the GW, we plot in in FIG. \ref{OmegaPlot} the energy density $\Omega_{GW}(f,\eta)$ defined in (\ref{Omega}) (where the frequnce relates to the wave-vector as $k=\frac{2\pi f}{c}$). It is clear that  $\Omega_{GW}(k,\eta)$ varies rapidly with $f$ when $M_Y>H^\star$.  Superimposed on the $\Omega$-$f$ curve the energy density sensitivities of the  KAGRA upgrade \cite{Kagra37}, Advanced Virgo \cite{LIGODOC87}, and the Einstein Telescope \cite{ET Sensitivity}. It is clear that the geometric Proca enters the observational window for $M_Y\geq 2 H^\star$. In this sense, future upgrades to GW detectors should be sensitive to such fluctuations provided that $M_Y\gtrsim H^\star$.

At the end of this section, we can conclude our analysis as follows, we study the field in light of cosmological perturbation theory and in this context the field does not influence the scalar nor vector modes of perturbation. Hence it can be probed by its effect on tensor modes.In FIG. \ref{kmodes1} we presented that subhorizon modes dominate the power spectrum which implies SGWB is detectable by the available ground-based detectors. Moreover, in FIG.\ref{OmegaPlot} the geometric Proca signatures get pronounced at larger field mass and the GW energy density (which is a function of frequency and geometric Proca mass) becomes large enough to be detectable by the next generation of the GW detectors.

\section{Conclusion \label{conclusion}}

In this work, we have analysed the geometric Proca field's imprints on gravitational waves. Our model, the extended Palatini gravity, is based on metric-affine gravity theories with zero torsion and nonzero metricity. It is well-known that this model reduces to GR plus a geometric Proca field through non-metricity. This geometric Einstein-Proca system has the Proca field mass as a free parameter. In our analysis, we derived the tensor power spectrum and energy density for vector perturbations induced by the geometric Proca field. We then performed a detailed numerical analysis to determine the size and characteristics of the gravitational wave (GW) signal. 

Our analysis revealed that the geometric Proca field significantly enhances the gravitational wave signal. We found that this effect is more sensitive to the mass term so larger mass values the effect is more pronounced and the GW energy density can reach levels detectable by future upgrades of GW detectors into third-generation (3G). If these findings are confirmed through observation, it will provide evidence that gravity is non-Riemannian in nature and that its reduction to general relativity (GR) introduces a geometric Proca field as a non-gauge vector boson. On the other hand, a lack of observational evidence would hugely constraint this class of theories so it would be disfavored direction of research. This could be a asimilar scenario to scalar-tensor theories after GW170817 \cite{Sakstein:2017xjx}.

A confirmation on non-Riemannian gravity would have far-reaching implications for our understanding of the universe and its evolution, including shedding light on the origin of PGWs and the nature of dark matter. Further possible directions for this analysis include studying the effects of the geometric Proca field on other astrophysical phenomena, such as the formation of galaxies and the evolution of black holes. In addition to investigating the possibility of using the geometric Proca field as a probe of the early universe.

\section*{Acknowledgements}
The work of S. K. is partially supported by Science, Technology $\&$ Innovation Funding Authority (STDF) under grant number 37272. DD is grateful to Elham Ghorani and Beyhan Puli{\c c}e for fruitful discussions on the geometric Proca field.

\bibliographystyle{plain}





\end{document}